\documentclass[amsmath,amssymb,pra,aps,showpacs,twocolumn,10pt,superscriptaddress]{revtex4-2}
\usepackage[colorlinks=true,citecolor=blue,linkcolor=red,urlcolor=blue]{hyperref}
\usepackage{graphicx}
\usepackage{dcolumn}
\usepackage{bm}
\usepackage[utf8]{inputenc}
\usepackage[T1]{fontenc}
\usepackage{lmodern}
\usepackage{cleveref}
\usepackage{color}
\usepackage{amsmath}
\usepackage{amssymb}
\usepackage{amsfonts}
\usepackage{latexsym}

\graphicspath{{./figs}}

\begin{document}

\title{Impact of the form of weighted networks on the quantum extreme reservoir computation}

\author{Aoi Hayashi}  \email{aoi.hayashi@oist.jp}
\affiliation{School of Multidisciplinary Science, Department of Informatics, SOKENDAI (the Graduate University for Advanced Studies), 2-1-2 Hitotsubashi, Chiyoda-ku, Tokyo 101-8430, Japan}
\affiliation{Okinawa Institute of Science and Technology Graduate University, Onna-son, Okinawa 904-0495, Japan}
\affiliation{National Institute of Informatics, 2-1-2 Hitotsubashi, Chiyoda-ku, Tokyo 101-8430, Japan}

\author{Akitada Sakurai}
\affiliation{Okinawa Institute of Science and Technology Graduate University, Onna-son, Okinawa 904-0495, Japan}
\affiliation{National Institute of Informatics, 2-1-2 Hitotsubashi, Chiyoda-ku, Tokyo 101-8430, Japan}

\author{Shin Nishio}
\affiliation{School of Multidisciplinary Science, Department of Informatics, SOKENDAI (the Graduate University for Advanced Studies), 2-1-2 Hitotsubashi, Chiyoda-ku, Tokyo 101-8430, Japan}
\affiliation{Okinawa Institute of Science and Technology Graduate University, Onna-son, Okinawa 904-0495, Japan}
\affiliation{National Institute of Informatics, 2-1-2 Hitotsubashi, Chiyoda-ku, Tokyo 101-8430, Japan}

 \author{William J. Munro}
 \affiliation{Okinawa Institute of Science and Technology Graduate University, Onna-son, Okinawa 904-0495, Japan}
 \affiliation{NTT Basic Research Laboratories \& Research Center for Theoretical Quantum Physics,  3-1 Morinosato-Wakamiya, Atsugi, Kanagawa, 243-0198, Japan}
 \affiliation{National Institute of Informatics, 2-1-2 Hitotsubashi, Chiyoda-ku, Tokyo 101-8430, Japan}
 
\author{Kae Nemoto}  \email{kae.nemoto@oist.jp} 
\affiliation{Okinawa Institute of Science and Technology Graduate University, Onna-son, Okinawa 904-0495, Japan}
\affiliation{National Institute of Informatics, 2-1-2 Hitotsubashi, Chiyoda-ku, Tokyo 101-8430, Japan}
\affiliation{School of Multidisciplinary Science, Department of Informatics, SOKENDAI (the Graduate University for Advanced Studies), 2-1-2 Hitotsubashi, Chiyoda-ku, Tokyo 101-8430, Japan}


\begin{abstract}
The quantum extreme reservoir computation (QERC) is a versatile quantum neural network model that combines the concepts of extreme machine learning with quantum reservoir computation.
Key to QERC is the generation of a complex quantum reservoir (feature space) that does not need to be optimized for different problem instances.
Originally, a periodically-driven system Hamiltonian dynamics was employed as the quantum feature map.
In this work we capture how the quantum feature map is generated as the number of time-steps of the dynamics increases by a method to characterize unitary matrices in the form of weighted networks.
Furthermore, to identify the key properties of the feature map that has sufficiently grown, we evaluate it with various weighted network models that could be used for the quantum reservoir in image classification situations.
At last, we show how a simple Hamiltonian model based on a disordered discrete time crystal with its simple implementation route provides nearly-optimal performance while removing the necessity of programming of the quantum processor gate by gate.

\end{abstract}

\maketitle

\section{Introduction}
In recent years we have seen the steady growth of the number of qubits available on a variety of quantum processors \cite{Arute2019, Friis2018, Zhong2020, Madsen2022}.  This has led to the new phase of quantum computer development, often called the ``NISQ'' era.  Here NISQ stands for noisy intermediate-scale quantum, which indicates that the quantum processor is too small to implement logical quantum operations and hence is inherently noisy.  The number of qubits in these quantum processors (well in excess of 50 \cite{Zhong2020, Gong2021, IBMEagle}) has already reached the point where the quantum computational tasks they can perform are intractable in a conventional computer, however, noise prevents us to extract the quantum advantage such quantum computer promise.  Hence, for the NISQ era to mark its significance in computer history, quantum advantages for real applications have to be demonstrated.

Many of the current NISQ processors are designed to operate via quantum gates \cite{Arute2019, Wright2019, Frey2022, Bharti2022}. To run a quantum algorithm, we need to obtain a quantum gate circuit from the quantum algorithm and then decompose each quantum gate into one implementable on the quantum processor at hand.  The noise in these quantum processors necessitates the optimization of quantum gate circuits to minimize their effect.  As long as the physical qubits are directly used for computation, quantum algorithms also need to be relatively short and resilient to noise.  Variational quantum algorithms (VQAs) have attracted a lot of attention from this viewpoint and have been intensively investigated \cite{Havlicek2019, Schuld2019, Noori2020}. However, there have been several issues with them, with the most significant obstacle being the difficulty in the optimization of the variational models \cite{McClean2018, Bittel2021}.  VQAs are a type of model of quantum neural networks (QNNs). It is well known that there are other models for using QNNs.  One such example is quantum reservoir computation \cite{Fujii2017, Ghosh2019, Pena2021, Bravo2022, Mujal2021}, which should be expected to be more implementation friendly.  Similarly to the chaotic dynamics used in the (classical) reservoir computation \cite{Bertschinger2004, Boedecker2012}, the quantum reservoir is to generate complex dynamics in the quantum system.  Realizing such dynamics using a quantum gate circuit approach is, however, not that simple \cite{Weinstein2008, Harrow2009, Neill2018, Boixo2018, Arute2019}. We do not require precise programming to generate a sufficient complexity in the quantum reservoir to realize our quantum algorithm\cite{Ghosh2019, Bravo2022}.  Instead, an effective quantum reservoir can potentially be generated by a simpler quantum system giving us a better way to utilize the computational power of QNNs.

Recently the quantum extreme reservoir computation (QERC) was proposed \cite{Sakurai2022} as a more advanced yet simpler QNN model based on reservoir computation and extreme machine learning \cite{Huang2006}.  This model uses a quantum reservoir to generate a quantum neural network which is then used for extreme machine learning.
The use of quantum reservoirs for extreme machine learning has been discussed.
However, there had been no attempts to image classifications until recent years \cite{Mujal2021}, and the previous models have been presented only in a general form.
Unlike the previous models, the QERC was the first concrete one that can perform the MNIST image classification task, which is considered an important task in computer vision.
This model has numerically shown to achieve the highest accuracy in classifying handwritten digits using the MNIST dataset with the smallest number of qubits \cite{Martyn2020}.
An interesting feature of this approach is that it utilizes a discrete time crystal (DTC) as the feature map, which is much simpler to implement than the quantum gate circuit needed to generate a random unitary matrix. This suggests that if we could understand the mechanism associated with using the complexity of the quantum dynamics for generating an effective feature space, it would become possible to design quantum feature maps more efficiently.

One of the versatile methods to study the complexity of quantum dynamics is to characterize it as a complex network.
Such network approaches have allowed us to quantify the complexity of quantum states around critical points of quantum phase transitions \cite{Andrew2017}, to build a graphical calculus for Gaussian pure states \cite{Nicolas2011}, and to reveal the preferential attachment mechanism during the melting process of the DTC \cite{Estarellas2020}.
Further these network approaches can be also applied to analysis of quantum machine learning models.
In Ref.~\cite{Sakurai2022}, the performance behavior of the QERC with the DTC dynamics was explored with the complex network emerging in the Hilbert space of the DTC dynamics \cite{Estarellas2020}.

Now let us outline the focus and structure of this paper. We will use the QERC proposed in \cite{Sakurai2022} as a tool to investigate the role of the feature map in quantum neural networks.  This model provides a convenient platform to do so, as the quantum contribution fully relies on the quantum reservoir.  We start in Section \ref{sectionII} with a description of the QERC model.
In Section \ref{sectionII}, we also present our method to characterize the unitary map, which is responsible for the quantum reservoir, as a complex network.
By our method, we investigate the feature map properties with the DTC and some random unitaries and will observe the difference in their dynamics in Section \ref{sectionIII}. Then, in Section \ref{sectionIV}, by benchmarking the QERC performance with those unitaries with concrete practical tasks, we discuss what properties of the performance arises based on the difference in the models' dynamics. We will confirm that the difference is not only the quantum reservoir properties, but also, in fact, can be exploited for a better QERC performance for the practical tasks. In Section \ref{sectionV}, we summarize our results.

\section{the QERC model and its characterization}
\label{sectionII}

\begin{figure*}[t]
\centering
\includegraphics[width=12.5cm]{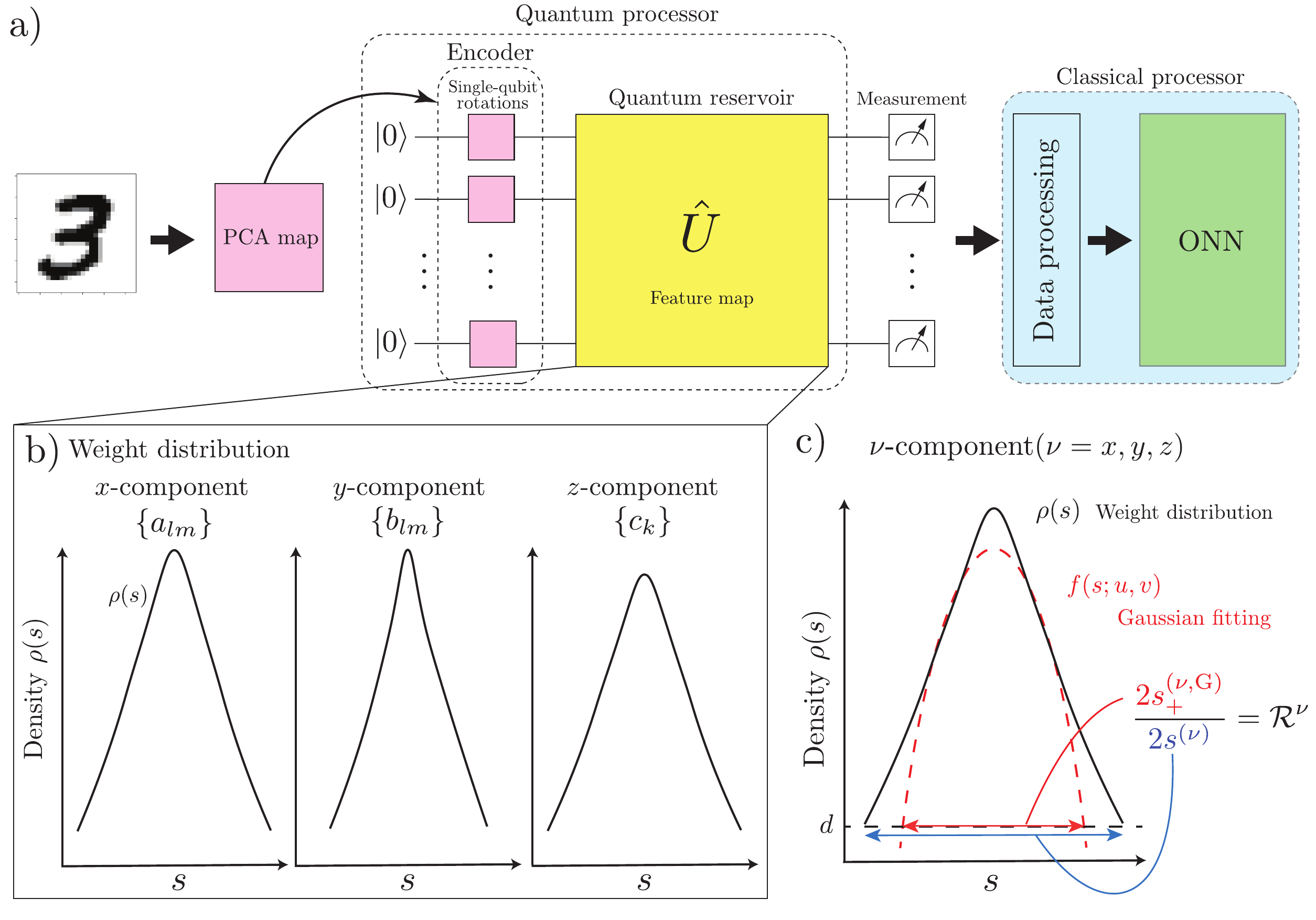}
\caption{a) Schematic architecture of the quantum extreme reservoir computation (QERC) processor. It begins with an image of size $28 \times 28$ pixels, is processed through principal component analysis, and is compressed to $2L$ components (where $L$ is the number of qubits). Using these $2L$ components, an initial state corresponding to the image is created by single-qubit rotations. The quantum reservoir then lets the initial state evolve. By projective measurements on the computational basis, the final state is converted to classical information. The amplitude distribution of this classical information is fed into the one-layer neural network (ONN). b) Schematics of the weight distributions for the $x$-, $y$-, and $z$-components. The panel c) summarizes the definition of the ratio $\mathcal{R}^\nu = s^{(\nu, \text{G})}_+ / s^{(\nu)}$ for the $\nu$-component, where $s^{(\nu)}$ is given by $s^{(\nu)} = (s_\text{max}^{(\nu)} - s_\text{min}^{(\nu)})/2$ with $s_\text{max (min)}^{(\nu)}$ being the maximum (minimum) value of $s$ in the $\nu$-component. Next one can consider a Gaussian fitting function $f(s; u, v) = u\exp(-vs^2)$, which allows us to define $s^{(\nu, \text{G})}_+ := \sqrt{\ln(u / d) / v}$ where $d$ corresponds to the possible minimum nonzero height of the density function $\rho(s) = h(s) / N_\nu ds$, that is, $d = (N_\nu ds)^{-1}$.}
\label{model_fig}
\end{figure*}

Let us begin with a brief description of QERC.  As shown in Fig. \ref{model_fig}a), QERC can be described in terms of three key components: the encoder, the quantum reservoir, and the classical processor: 

\begin{description}
\item[\emph{Encoder}] Here, the data to be classified is preprocessed (if necessary) and encoded into the initial state of the quantum reservoir.  In more detail, as shown in Fig. \ref{model_fig}a), a Principal Component Analysis (PCA) map is used for the preprocessing of the classical data. Then an appropriate encoding strategy needs to be chosen for the problem at hand.  For a quantum reservoir of $L$ qubits, the $2L$ most significant parameters from the PCA map will be encoded by single-qubit rotations. 
 
\item[\emph{Quantum reservoir}] In this step, the quantum reservoir provides the feature space for QERC.  The quantum dynamics of the quantum reservoir determines the feature-map properties for the quantum computation, which is given by the unitary operator $\hat{U}$.

\item[\emph{Classical processor}]  In this final step, the state given by the unitary operator $\hat{U}$ acting on the initial state is measured projectively on the computational basis.  The process will be repeated to obtain the amplitude distribution of the state generated by the unitary operator.  This amplitude distribution is then processed through a one-layer neural network (ONN). 
\end{description}
We immediately notice that this is a hybrid quantum-classical algorithm. In QERC the feature space is provided by the quantum reservoir, whereas the optimization is carried out on the classical processor (ONN). Typically the quantum reservoir does not need to be optimized for different problem instances \cite{Sakurai2022, Fujii2017, Bravo2022}.

Now our interest in this paper is the properties we require for the quantum reservoir and their influence on the performance of QERC. In particular, we want to show that how we set the quantum reservoir is important.  In this work, we first employ the DTC model used in \cite{Sakurai2022} as our choice of a quantum reservoir.  The DTC model has a parameter that controls the complexity of the dynamics, namely starting with the perfect discrete time crystal when the rotation parameter error $\epsilon=0$, the dynamics gradually deviates from a DTC, acquiring its complexity as $\epsilon$ increases.  This parameter $\epsilon$ represents an error in the single-qubit rotation in the DTC Hamiltonian, which is given by
\begin{equation}\label{DTC_Hamiltonian}
\hat{H}(t) = 
\begin{cases}
	\hat{H}_1 = \hbar g(1 - \epsilon)\sum_{l}\hat{\sigma}^x_l & t \in [0, T/2)\\
	\hat{H}_2 = \hbar\sum_{lm}J_{lm}\hat{\sigma}^z_l\hat{\sigma}^z_m + \hbar\sum_{l}D_l\hat{\sigma}^z_l & t \in [T/2, T)
\end{cases}
\end{equation}
where $\hat{\sigma}^a_l \ (a = x, y, z)$ represent the Pauli operators on the $l$-th qubit.  Next, $T$ is the cycle of driving, while the DTC cycle is $2T$.  Further $g$ is the rotation strength, and in this case, we set $gT = \pi$. Now $J_{lm} = J_0 / \lvert l - m\rvert^\alpha$ is the coupling strength between the qubits $l$ and $m$ with a power-law decay that scales with a constant $\alpha$. Finally, $D_l$ is a disordered external field for each qubit $l$. Unless explicitly stated, all the $D_lT$ are set to zero in this work.

The time-periodic system is conveniently characterized by the Floquet operator $\hat{\mathcal{F}} = \exp\left[-i\hat{H}_2T/2\hbar\right] \exp\left[-i\hat{H}_1T/2\hbar\right]$ where the stroboscopic time-evolution can be obtained by the unitary operator $\hat{U}(nT, 0) = \hat{\mathcal{F}}^n$ for $n \in \mathbb{N}$. Hence, we use the unitary operator $\hat{U}(nT, 0)$ for different values of $n$ to characterize the quantum reservoir.

\subsection{Characterization of the unitary matrices}

The next step is the characterization of the unitary operator $\hat{U}(nT, 0)$.
This unitary operator acts as a map between the input and the output states given by the feature map used for the quantum computation. Such a map can be considered as a weighted network \cite{Altaisky2001, Tacchino2019}. However as the unitary operators are defined on the complex field, the translation to a weighted network is not trivial.  Here we apply a generator decomposition of a unitary matrix $U \in $ U($N$), where $N$ is the dimension of the unitary matrix, in order to represent the unitary operator as a weighted network.  The generators of U($N$) are the Hermitian matrices forming the Lie algebra.

Now a unitary matrix $U \in $ U($N$) can be written in the form $U = e^{-iG}$ where $G$ is a Hermitian matrix. This Hermitian matrix can be represented with real coefficients $a_{lm}, b_{lm}$, and $c_k$ by the decomposition of $G$ with respect to the generators $\lambda$ as
\begin{equation}
G = \sum_{l < m} (a_{lm}\lambda^x_{lm} + b_{lm}\lambda^y_{lm}) + \sum_k c_k\lambda^z_k, \label{decomp_G}
\end{equation}
where those $\lambda$ generators are the generalized Gell-Mann matrices \cite{Greiner1994, Nemoto2000, Kimura2003}
\begin{eqnarray}
    (\lambda^{\nu}_{lm})_{ij} &=& \delta_{li}\delta_{mj}\sigma^{\nu}_{12} + \delta_{lj}\delta_{mi}\sigma^{\nu}_{21}\ \ \ (\nu = x, y), \\
    \lambda^z_{k} &=& \sqrt{\frac{2}{k(k + 1)}}~\text{diag}(\underbrace{1, \cdots, 1}_{k~\text{times}}, -k, 0, \cdots, 0) \\
     \lambda^z_N &=& I.
\end{eqnarray}
Here $\delta_{ij}$ represents the Kronecker delta, $\sigma^{\nu}_{ij}$ ($\nu = x, y$) is the ($i, j$)-component of the Pauli matrices and $I$ the identity matrix. Next $G$ as a weight matrix has three components $x, y,$ and $z$ with $\{a_{lm}\}, \{b_{lm}\}$ and $\{c_k\}$ being the $x, y$ and $z$ contributions of the weight matrix, respectively.

The Hermitian matrix $G$ obtained from the unitary matrix $U$ is not necessarily unique.  To uniquely determine $G$ for a given unitary matrix, we employ the  \emph{principal logarithm} of a matrix \cite{Higham2008} in our numerical analysis.  If $A$ is a complex-valued matrix of dimension $N$ with no eigenvalues on the negative real line $\mathbb{R}^-$, then there is a unique logarithm $X$ of a matrix $A$ such that all of its eigenvalues lie in the strip $\{z : -\pi < \text{Im}(z) < \pi\}$.   Here $X$ is called the \emph{principal logarithm} of $A$ and denoted by $X = \log(A)$.  
For the DTC model, we compute the Hermitian matrix for period $n$, $G(n) = i\log(\hat{U}(nT, 0))$ noting that $G(n)$ is not simply equal to $n \times G(1)$.

To convert the weight matrix to its weight distribution, we first count how many coefficients are in a certain value window $(s, s + ds]$ for $s, ds \in \mathbb{R}$.
This gives us a histogram $h(s)$ to show how likely the coefficients are to take a certain value $(s, s + ds]$.
In numerical calculations, we take 100 segments for each coefficient set to determine the value of $ds$:
let the support of the histogram denoted by a sequence $S = \{s_0, s_1, \cdots, s_{M-1}\}$, where $s_i - s_{i-1} = ds > 0$ for all $i = 0, 1, \cdots, M-1$ and $M$ is the number of segments in the range $(s_0, s_{M-1}]$.
Then, once we define $s_0$, $s_{M-1}$ and $M$, we have $ds = (s_{M-1} - s_0)/M$.
After obtaining the histogram, we have a density function $\rho(s) = h(s) / N_\nu ds$ where $N_\nu$ is the number of elements in the component $\nu = x, y$ or $z$, that is, $N_\nu = \sum_{i=0}^{M-2}h(s_i)$. We refer to this as the weight distribution of the $\nu$-components ($\nu = x, y, z$) (see Fig.~\ref{model_fig}b)).

\subsection{Characterization of the weight distributions}\label{section_ER}
We will show weight distributions for the DTC model in different configurations and other models in Sec.~\ref{sectionIII}.
To quantitatively characterize those weight distributions, we calculate two quantities for each weight distribution.
The first is the empirical standard deviation $\sigma_\nu$ is given by the standard deviation of values of elements in a component $\nu$, for example, in the case of $\nu = x$, $\sigma_x = \sqrt{\mathrm{var}_{lm}(a_{lm})}$.

Our second quantity is a ratio that represents how far the weight distribution reaches from its center compared with a Gaussian function approximating the weight distribution, which is analogous to the MP rank defined in Ref.~\cite{Mahoney2019}.
The ratio $\mathcal{R}^\nu$ for the weight distribution of the $\nu$-component is given by
\begin{equation}
\mathcal{R}^\nu = \frac{s^{(\nu, \text{G})}_+}{s^{(\nu)}}.
\end{equation}
The denominator $s^{(\nu)}$ is defined as a quantity representing how far the weight distribution reaches from its center in the horizontal axis.
Since the weight distribution $\rho(s)$ is not necessarily symmetric with respect to $s=0$, $s^{(\nu)}$ is given by $s^{(\nu)} = (s_\text{max}^{(\nu)} - s_\text{min}^{(\nu)})/2$ with $s_\text{max (min)}^{(\nu)}$ being the maximum (minimum) value of $s$ in the $\nu$-component, that is, $\mathrm{max}\{s \in S \vert \rho(s) \neq 0\}$ ($\mathrm{min}\{s \in S \vert \rho(s) \neq 0\}$).
Next, $s^{(\nu, \text{G})}_+$ represents how far a Gaussian distribution function reaches from $s=0$ obtained by a Gaussian fitting to the weight distribution.
The Gaussian function reaches $s=\infty$ in general.
Thus, we introduce a cutoff for the Gaussian function.
In more detail, let $f(s; u, v)$ denote the Gaussian function given by the Gaussian fitting with fitting parameters $u, v$: $f(s; u, v) = u\exp(-vs^2)$.
To introduce the cutoff, we consider cross points between $f(s; u, v)$ and a horizontal line at a value of $d$, that is, $d = f(s; u, v)$.
Then, $s^{(\nu, \text{G})}_+$ is given as $s^{(\nu, \text{G})}_+ = \sqrt{\ln(u / d) / v}$.
In our numerical calculations, we set $d = (N_\nu ds)^{-1}$, which corresponds to the possible minimum nonzero height of the density function $\rho(s) = h(s) / N_\nu ds$.
In Fig.~\ref{model_fig}c), the definition of $\mathcal{R}^\nu$ is summarized.

\subsection{Simulation setup for the QERC}

We begin our considerations here by first directly evaluating the properties of the feature map generated by our DTC dynamics using the method outlined above.  Setting a computational task is not essential to do the analysis, however, it is extremely useful when later we compare these properties to the performance of the QERC.  It is convenient to set a computational task to evaluate both at the same time and on a similar footing. In this paper, we use the well-known MNIST dataset \cite{MNIST} where each image has $784\ (= 28 \times 28)$ pixels. We employ PCA to reduce each image data to the $2L$ components, which can then be encoded in the initial state of the quantum reservoir of $L$ qubits by single-qubit rotations. Finally, to optimize the parameters of the ONN, we employ the stochastic gradient descent method used in \cite{Sakurai2022}.  Throughout this work, our parameters are set as $L = 10$, $ J_0T = 0.12$ with $\alpha = 1.51$, These are compatible with the current ion trap experiments\cite{Zhang2017}.  We also set $\epsilon=0.03$ as the highest accuracy rate that has been reported for the QERC with this parameter value \cite{Sakurai2022}.

\section{Quantum reservoir weight distribution}
\label{sectionIII}

\begin{figure}[bt]
\centering
\includegraphics[width=8.5cm]{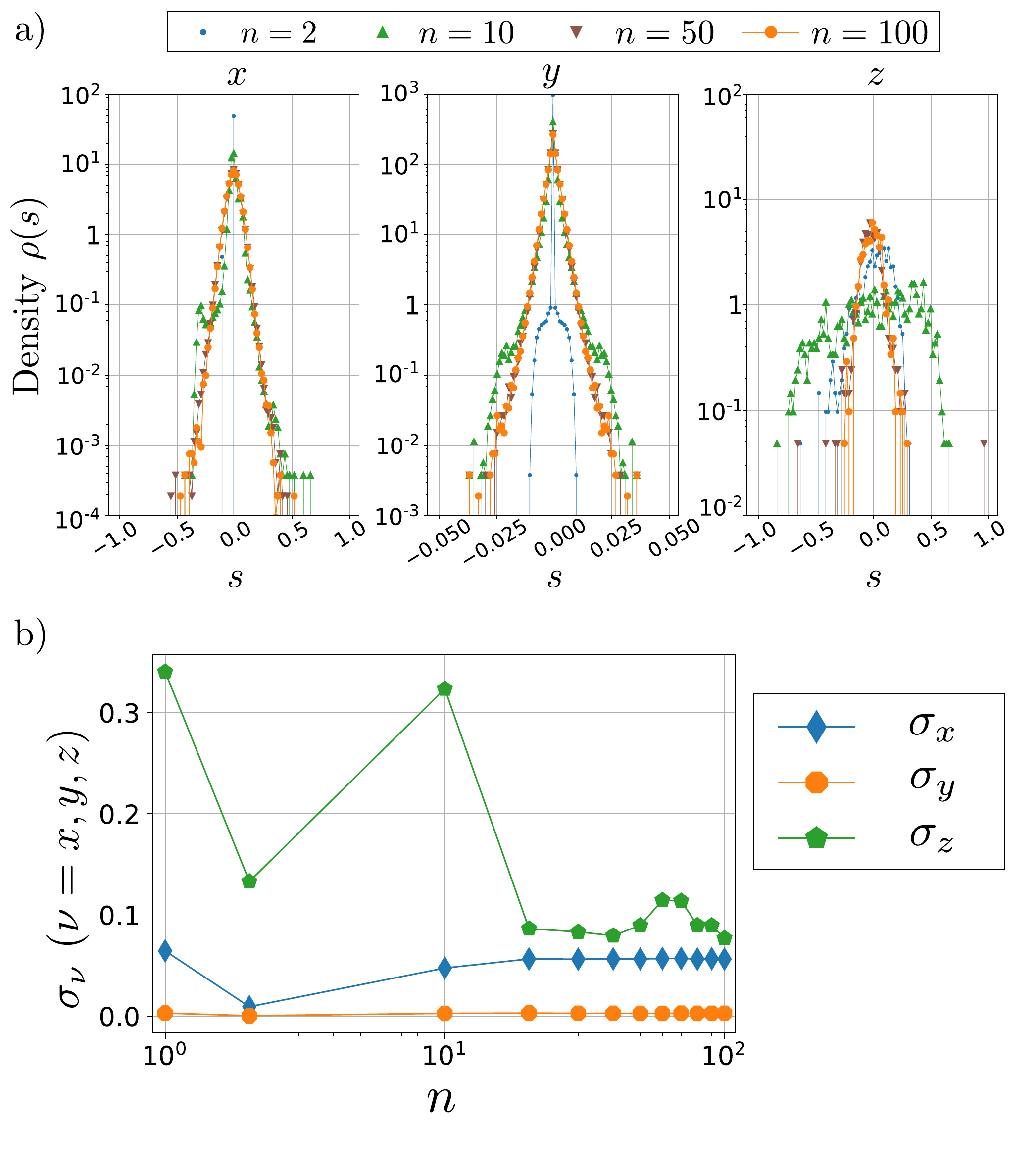}
\caption{a) Convergence of the weight distribution in the DTC model. From the left to right panels, the weight distribution functions for $a_{lm}, b_{lm}$ and $c_k$ are depicted where the colors correspond to different periods. the blue (dot), green (upward triangle), brown (downward triangle), and orange (filled circle) curves are for $n = 2, 10, 50$, and 100, respectively. b) The $n$-dependency of the empirical standard deviations, $\sigma_x = \sqrt{\mathrm{var}_{lm}(a_{lm})}$ (blue with diamonds), $\sigma_y = \sqrt{\mathrm{var}_{lm}(b_{lm})}$ (orange with hexagons), and $\sigma_z = \sqrt{\mathrm{var}_{k}(c_{k})}$ (green with pentagons).}
\label{DTC_dist}
\end{figure}

It is important to emphasize that unless perfectly periodical, the quantum dynamics of the DTC model deviates from its initial state in time as it evolves. This allows for the growth of the complexity in the system. To observe such complexity growth in the unitary dynamics, we evaluate the weight distribution of $G(n)$ for various time periods: $n = 2, 10, 50$, and 100. From Eq.\ref{decomp_G}, we can determine the real coefficients $a_{lm}, b_{lm}$ and $c_k$ characterizing $G(n)$ for each $n$.
The Hermitian weight matrix $G(n)$ is equivalent to the $n$-period effective Hamiltonian up to the constant factor $\hbar/nT$. Thus, the diagonal (corresponding to $\{c_k\}$) and off-diagonal (corresponding to $\{a_{lm}\}, \{b_{lm}\}$) entries of $G(n)$ are associated with the energies of the basis states and the transition energies between the basis states, respectively.

In Fig ~\ref{DTC_dist}a) we plot the weight distribution for the $x$ ($a_{lm}$), $y$ ($b_{lm}$) and $z$ ($c_k$) components of $G(n)$ respectively.  
Each color (symbol) represents a different period of the time evolution.
For $n=2$ (blue line), we observe very sharp peaks at $s = 0$ for the $x$ and $y$ components.  As these components correspond to the off-diagonal entries of the Hermitian matrix $G(n)$, the sharp peaks around $s = 0$ mean very few transitions between the basis states for this time period.
However, for large periods, $n = 50$ (brown curve), 100 (orange curve), the weight distributions for all the components converge to a similar shape that is approximately quadratic in the log-scaled plots (gaussian in linear plot).
In the middle of these two time regions, at $n=10$ (green curve) the $x$ and $y$ components have already converged to the typical distribution, however, the $z$ component has broadened the most.
This suggests that there is a tradeoff in this time regime; the stationary elements of the $z$ component significantly suppress the effect of the $x$ and $y$ components.
This trade-off captures the dynamics of the DTC melting slowly in time in this ($\epsilon=0.03$) parameter regime.

The behavior for those components can be quantitatively observed by the empirical standard deviation for the weight distributions, defined in Sec.~\ref{section_ER}.
In Fig.~\ref{DTC_dist}b), the empirical standard deviations are depicted against the number of periods, $n$.
While the $z$-component is broadened at $n=10$, the $x$- and $y$-components gradually get higher standard deviations, and then they converge.

\subsection{Comparing the weight distribution}\label{Sec_HandC}

To capture the characteristics of the weight distribution for the DTC model, we first introduce the Haar measure sampling of unitary operators \cite{Emerson2005, Boixo2018, Mullane2020}.  The Haar measure sampling can be considered to exhibit a typical complexity that a quantum computer may provide, and its gate implementation is usually given through unitary t-design \cite{Weinstein2008, Harrow2009}.  Hence the similarity and disparity in the weight functions for these cases would give us valuable insights into understanding the DTC dynamics and its role in QERC. In our analysis, to obtain a typical distribution given by the Haar measure sampling, the $N \times N$ unitary matrix $U_\text{H}$ is created using the QR decomposition \cite{Mezzadri2007} where $N = 2^L =2^{10}= 1024$. We compare this unitary map, which we refer to as the Haar-random model, to the converged weight distribution of the DTC model.

\begin{figure}[t]
\centering
\includegraphics[width=1\linewidth]{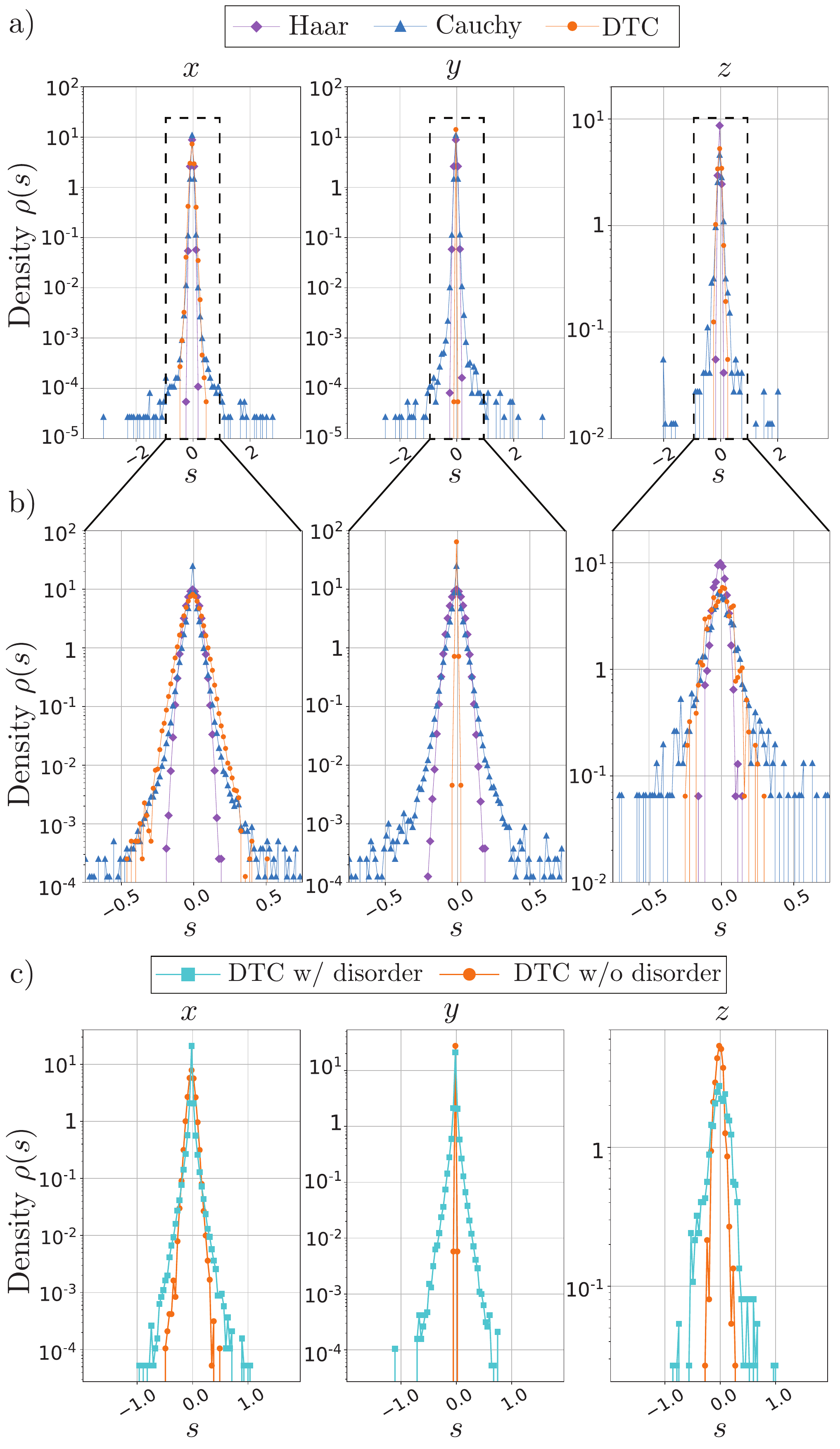}
\caption{Comparison of the Haar-random, Cauchy-random, and DTC models for $n = 100$ in a). In each case from the left to right panels, the distributions of the $x$-, $y$-, and $z$-components are depicted respectively. In b), the weight distributions of those models are shown for the range $[-0.75, 0.75]$. Finally, in c), a comparison is shown between the DTC model with and without disorder for $n = 100$.}
\label{Comparison_distributions}
\end{figure}

Now as shown in Fig.~\ref{Comparison_distributions}a) and b), the typical weight distributions are approximately Gaussian for all components $x, y$, and z.  Here we use only one sample from the Haar-random model since one sample and not the average of many samples will be used within the QERC. Further, we do not lose generality as discussed in Appendix.~\ref{appendix_randomness}.

We can now compare the DTC model to the Haar-random model, where we characterize the weight distributions of the DTC model with two properties, broadness, and tail.
Although the DTC's $y$-component has a narrower distribution compared to the Haar-random model, the broadness of the distribution for the $x$- and $z$- components are comparable as shown by the empirical standard deviations in Table.~\ref{sigma_R_table}.

However, only the DTC model has a tail in the weight distribution of the $x$ component; a few large elements at the edge of the weight distribution.
To quantitatively observe tails in the weight distribution, we calculate the ratio $\mathcal{R}$ defined in Sec.~\ref{section_ER}.
Table.~\ref{sigma_R_table} shows the averages of the ratios for the DTC and Haar models where $\overline{\mathcal{R}}$ denotes the average, that is, $\sum_{\nu = x, y, z}\mathcal{R}^\nu / 3$.
One can see that the averaged ratio of the DTC model is smaller than that of the Haar model, which is close to the unit.
It implies that the weight distribution of the DTC model deviates from that of the Haar model in terms of the tail.

\begin{table}[t]
\centering
\begin{tabular}{|c||c|c|c|c|}
\hline
 & $\sigma_x$ & $\sigma_y$ & $\sigma_z$ & $\overline{\mathcal{R}}$ \\ \hline\hline
Haar & 0.0400 & 0.0402 & 0.0406 & 0.8592 \\ \hline
Cauchy & 0.0402 & 0.0394 &  0.2733 & 0.0978 \\ \hline
DTC & 0.0564 & 0.0028 & 0.0770 & 0.5189 \\ \hline
DDTC & 0.0410 &  0.0393 & 0.1914 & 0.2245 \\ \hline
\end{tabular}
\caption{The empirical standard deviations and the ratios $\mathcal{R}$ for different models. $\overline{\mathcal{R}}$ denotes the average of the ratios over three components, that is, $\overline{\mathcal{R}} = \sum_{\nu=x, y, z}\mathcal{R}^\nu / 3$.}\label{sigma_R_table}
\end{table}

Next to explore the difference associated with the tail we found in the DTC models distribution, we employ the Cauchy distribution.  The reason for this is as follows. In classical reservoir computation, the Cauchy distribution was used to obtain the edge of chaos where the reservoir computation should be optimal \cite{Toyoizumi2020}.  Hence it is interesting to see the properties of the feature map generated by the Cauchy distribution.   The Cauchy distribution is given by
\begin{equation}
\text{Cauchy}(x; \gamma) =\left( \frac{1}{\pi\gamma}\right)\frac{1}{1 + (x / \gamma)^2},\label{def_Cauchy}
\end{equation}
where $\gamma$ is the scale parameter. Since the Cauchy distribution has a power-law tail, one would expect that the weight distribution exhibits a long tail. The unitary matrix $U_\text{C}$ for this Cauchy-random model is defined as follows.
First, we generate an $N \times N$ Hermitian matrix $A$ whose real and imaginary parts in each independent entry are drawn from the Cauchy distribution~(\ref{def_Cauchy}). Then we define the unitary matrix $U_\text{C}$ as $U_\text{C} = e^{-i A}$.

Further, we set $\gamma = 0.04$ in Eq.~\ref{def_Cauchy} for consistency with the corresponding parameter of the Haar-random model ($\sigma \approx 0.04$), and the size $N$ is set as $N = 2^L = 1024$.

Applying the decomposition~(\ref{decomp_G}) to $G_\text{C} = i\log(U_\text{C})$, we obtain each weight distribution for the three components $x, y$ and $z$., which are depicted in Fig.~\ref{Comparison_distributions}a) and b). The weight distributions for the Cauchy-random model definitely have a tail, much longer than that in the DTC model, for all the components $x$, $y$, and $z$.
Table.~\ref{sigma_R_table} shows the averaged ratio of the Cauchy-random model, which is even smaller than the other two models.
This reflects the nature of the Cauchy distribution~(\ref{def_Cauchy}), and we will come back to this point later.

\section{Relation between the QERC performance and the weight distribution}
\label{sectionIV}
\begin{figure}[t]
\centering
\includegraphics[width=8.0cm]{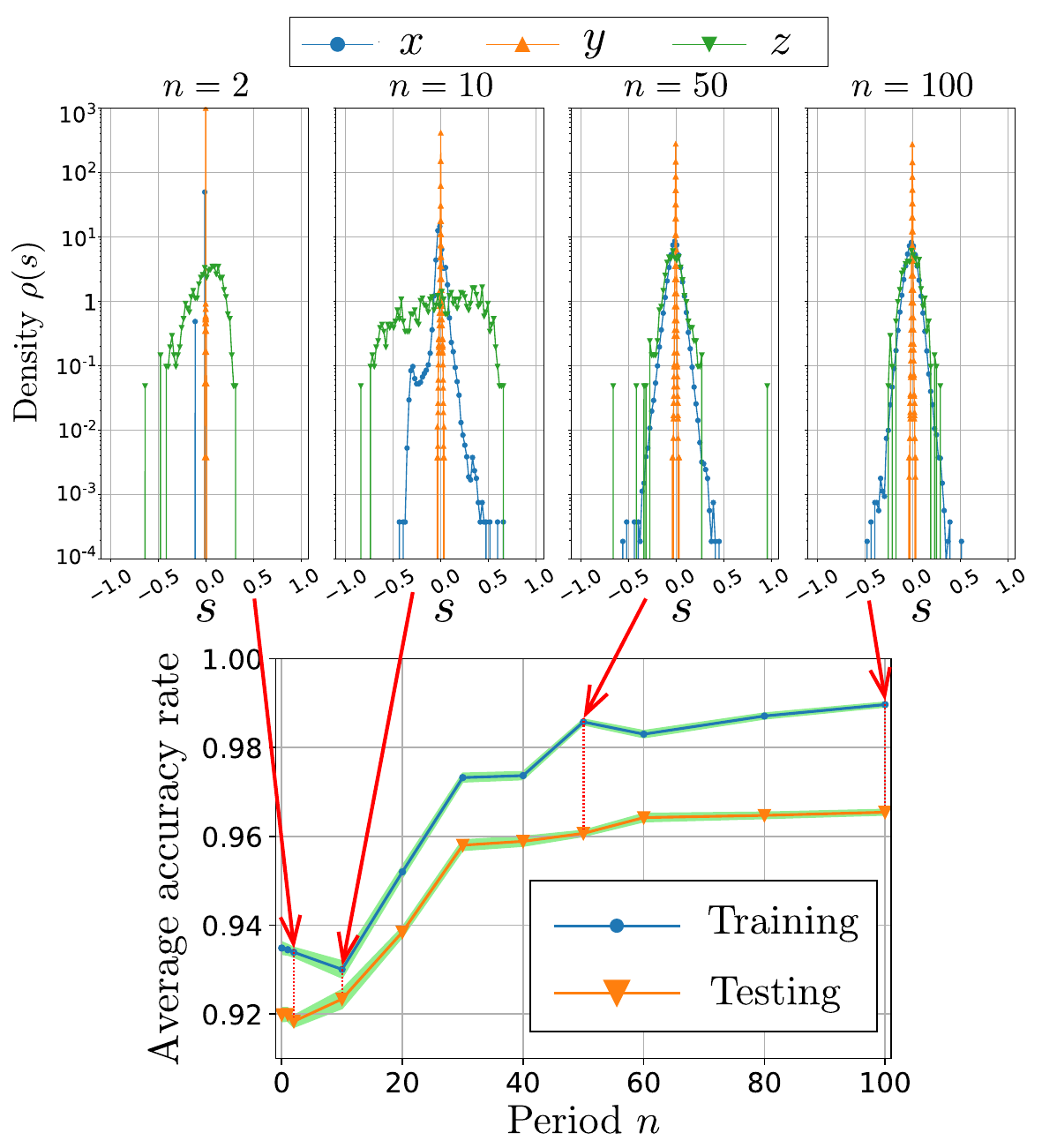}
\caption{Average accuracy rates for training and testing with the associated standard deviation against the period in the DTC model. The blue (dot) and orange (downward triangle) curves correspond to training and testing, respectively. At each datapoint, the average and the standard deviation are taken for 250 to 300 epochs in the ONN optimization.}\label{acc_vs_period}
\end{figure}

As we have characterized the three models through the weight distribution, let us now turn our attention to the performance of the QERC employing these three different models as its feature space.
In the previous work \cite{Sakurai2022} it was shown that the accuracy of the QERC increases with the number of the time periods of the DTC model saturating near $n=50$.  The behavior of this accuracy rate can be predicted from the time evolution of the weighted distributions seen in Fig.~\ref{DTC_dist}, as the unitary map of the DTC model acquires the typical complexity around $n=50$.  To illustrate this further, Fig.~\ref{acc_vs_period} summarizes the comparison between the accuracy rate and the weight distribution. Here we plot the accuracy rates for training (blue dot) and testing (orange downward triangle) against the time period $n$ in the DTC model and insert the weight distributions for $n = 2, 10, 50, 100$.

The broadness in the $x$- and $y$-components of the weight distribution are essential for the quantum reservoir to achieve a higher performance.  The trade-off between the $x, y$ components and $z$ component is reflected in the average accuracy rates.  This suggests that even if the system dynamics is complex enough, within a short coherent regime, the system does not evolve enough to achieve the computational power the system would promise.  

The complexity generated in a finite system has to be bounded, and unlike unitary maps from the Haar-measure sampling, the DTC model does not reach the maximum randomness allowed for the system to have certain tendencies in its dynamics.   Next, we further investigate the effect of this difference in these models on the performance of the QERC.

\subsection{Tails in the Distribution and the Performance}\label{Sec_tail}

The unitary operators we characterized through the weight distributions directly serve as the feature map for the QERC.  We will now explore how these different weight distributions and their associated feature maps affect the performance of the QERC with the MNIST dataset.

\begin{table}[t]
\centering
\begin{tabular}{|l||c|c|c|}
\hline
a) MNIST & testing acc. (std.) & training acc. (std.) & $\Delta_\text{acc.}$ \\ \hline\hline
PCA & $0.8635 (\pm 0.0009)$ & $0.8688 (\pm 0.0005)$ & 0.0053 \\ \hline
Haar & $0.9657(\pm 0.0005)$ & $0.9949(\pm 0.0003)$ & 0.0292\\ \hline
Cauchy & $0.9673(\pm 0.0005)$ & $0.9945(\pm 0.0003)$ & 0.0272\\ \hline
DTC & $0.9655(\pm 0.0006)$ & $0.9897(\pm 0.0005)$ & 0.0242 \\ \hline
DDTC & $0.9671(\pm 0.0005)$ & $0.9911(\pm 0.0005)$ & 0.0240\\ \hline\hline
\multicolumn{4}{|l|}{b) FashionMNIST}\\ \hline\hline
Haar & $0.9427(\pm 0.0017)$ & $0.9744(\pm 0.0017)$ & 0.0317\\ \hline
Cauchy & $0.9440(\pm 0.0029)$ & $0.9720(\pm 0.0023)$ & 0.0280 \\ \hline
DTC & $0.9388(\pm 0.0039)$ & $0.9662(\pm 0.0049)$ & 0.0274\\ \hline
DDTC & $0.9407(\pm 0.0050)$ & $0.9589(\pm 0.0044)$ & 0.0182\\ \hline
\end{tabular}
\caption{Average accuracy rates with the associated standard deviation of the various feature models with a) the MNIST and b) FashionMNIST datasets. For the FashionMNIST case, we picked three classes: T-shirt, Pullover, and Dress. The average and the standard deviation are taken from 250 to 300 epochs, and $\Delta_\text{acc.}$ denotes the gap between training and testing. The label ``PCA'' denotes the case where the PCA components are directly fed into the ONN (without any quantum feature maps). The results for the DTC cases with/without disorder are for the period $n = 100$. In the random models, the accuracy rates are from a specific realization. }\label{acc_table}
\end{table}

Table.~\ref{acc_table}a) presents the accuracy rates for each model (see also Appendix.~\ref{appendix_accplot}).

Before the comparison of the quantum feature maps, we first provide the performance of the case where the PCA components are directly fed into the ONN (without quantum feature maps).
One can immediately observe that the case, denoted by ``PCA'' in Table.~\ref{acc_table}a), has the lower accuracy rates in both training and testing than any other cases with quantum feature maps.
It states that those quantum feature maps significantly help the QERC achieving a high performance.

Next, we compare the accuracy rates of the quantum feature maps.
In testing, the DTC $(n=100)$ model is comparable to the Haar-random model, whereas not so in training.
Moreover, it is interesting to notice that the Haar-random model does not give the best testing accuracy rate in this setting, but the Cauchy-random model does.
These observations may suggest that the tail in the weight distribution of the DTC and Cauchy models contributes to the higher accuracy rate.

\begin{figure}[t]
\centering
\includegraphics[width=7cm]{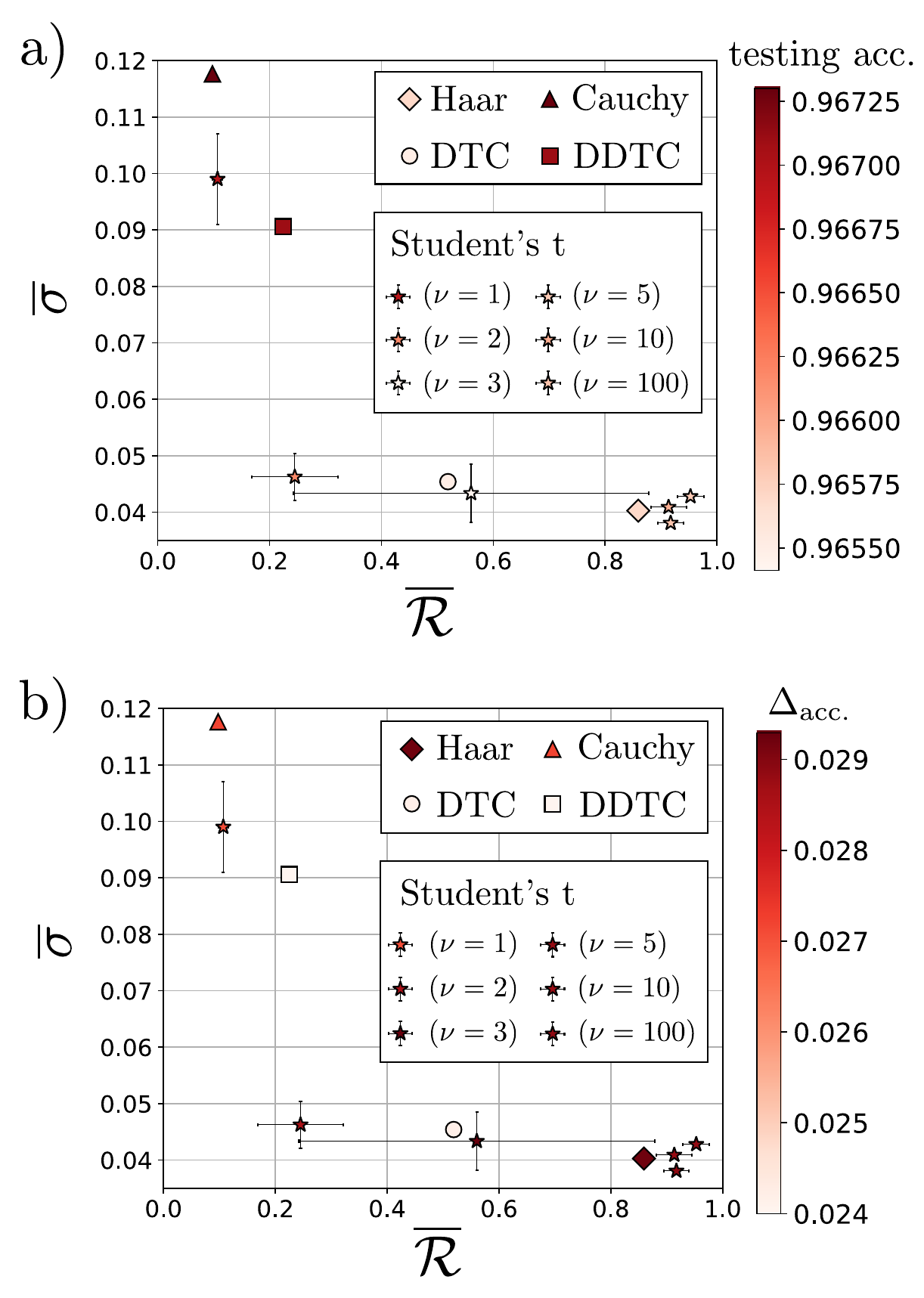}
\caption{Scatter plot of the models we consider in this paper against $\overline{\sigma}$ and $\overline{\mathcal{R}}$ with colored markers indicating a) the testing accuracy rate and b) the gap $\Delta_\text{acc.}$ of accuracy rates between training and testing.}\label{R_Sigma_acc}
\end{figure}

In order to see the relation between the accuracy rate and the weight distribution, we show the correlation between the testing accuracy rate and the quantities we used to characterize the weight distribution: the averaged empirical standard deviation ($\overline{\sigma}$) and ratio ($\overline{\mathcal{R}}$) in Fig.~\ref{R_Sigma_acc}a).
The quantities correspond to the vertical and horizontal axes, respectively, and the color of the markers indicates the testing accuracy rate.
One can find that the color becomes darker as the ratio $\overline{\mathcal{R}}$ gets smaller.

To investigate this further, we introduce the t-random model, whose unitary is defined using the Student's t-distribution in the same way to generate the Cauchy-random model.
The Student's t-distribution is defined as
\begin{equation}
\mathrm{t}(x; 0, \gamma_\mathrm{t}, \nu) = \frac{\Gamma(\frac{\nu+1}{2})}{\gamma_\mathrm{t}\sqrt{\pi\nu}\Gamma(\frac{\nu}{2})}\left(1 + \frac{(x/\gamma_\mathrm{t})^2}{\nu}\right)^{-(\nu+1)/2}
\end{equation}
where $\Gamma(\cdot)$ denotes the Gamma function, and $\gamma_\mathrm{t}$ is a scale parameter.
$\nu$ is a parameter determining how heavy the tail of the distribution is.
This parameter connects the standard Cauchy ($\nu=1$) and normal ($\nu=\infty$) distributions when $\gamma_\mathrm{t}=1$.
In our context, this parameter allows us to generate the weight distribution located in between the Haar- and Cauchy-r
random models in Fig.~\ref{R_Sigma_acc}a) with $\gamma_\mathrm{t} = \sigma = \gamma =  0.04$.

In Fig.~\ref{R_Sigma_acc}a), the stars correspond to the averaged data points over ten realizations of the unitary for each value of $\nu$ ($\nu = 1, 2, 3, 5, 10,$ and 100). The error bars correspond to the standard deviations of each data point (see Appendix.~\ref{appendix_t} for detailed results of the t-random model).
One can find that the ratio and the testing accuracy rate tend to be smaller and darker, respectively, as $\nu$ gets smaller.
Moreover, data points close to each other in the plot have similar testing accuracy rates.
This simulation with the Student's t-distribution illustrates the correlation between the testing accuracy rate and the tail in the weight distribution.

Considering the implementability of the QERC with an effective quantum feature map, one may wonder what physical system is close to the Cauchy- and t-random (for $\nu = 1$) models in Fig.~\ref{R_Sigma_acc}a) and achieves a high testing accuracy rate since the DTC model is far from such models, and there still seems to be room to improve it.
We here consider disorder to the DTC model in Eq.~\ref{DTC_Hamiltonian}, and actually, the disordered DTC (DDTC) model has a higher accuracy rate, as we will see later.

In Eq.~\ref{DTC_Hamiltonian}, we have set $D_lT = 0$ to enable us to consider the DTC model without the disorder. That constraint can now be relaxed. The disorder in Floquet systems has been considered to be important to suppress the thermalization and stabilize the DTCs \cite{Zhang2017, Frey2022}.  Actually, it is more realistic to have a little disorder in such quantum systems, and so it is worth checking if our QERC's performance is robust to such disorder.  We choose the disorder terms $D_lT$ in Eq.\ref{DTC_Hamiltonian} independently drawn from a uniform distribution on $[0, 2\pi)$.
As illustrated in Fig.\ref{Comparison_distributions}c), the introduction of the disorder changes the form of the weight distribution, and it is characterized by the empirical standard deviation and the ratio (see Table.~\ref{sigma_R_table}).
So the DDTC model is now located around the upper left region of Fig.~\ref{R_Sigma_acc}a), and one can find that the model achieves a similar testing accuracy rate to that of the Cauchy- and t-random (with $\nu = 1$) models (see also Table.~\ref{acc_table}a)).

So far, we have only discussed our testing accuracy, and it is important we now turn our attention to the training accuracy and the difference between the testing and training accuracy rates. The difference $\Delta_{\text{acc.}}$ is an important parameter in terms of the overfitting and the generalization performance of this machine learning model.  Neural networks often show the effects of overfitting \cite{Srivastava2014, Caruana2000} where the neural networks are too well optimized to the training date and lose their flexibility to deal with the testing data. The generalization performance is hence an important factor in designing QNNs.  

In Table.\ref{acc_table}, we also provide the training accuracy and difference $\Delta_{\text{acc.}}$ for the various feature models we have considered.
We also plot the correlation between the generalization performance and the properties of the weight distribution in Fig.~\ref{R_Sigma_acc}b).
One can observe that the Cauchy-random model has the smallest gap $\Delta_{\text{acc.}}$ among the artificial models, and the value of $\Delta_{\text{acc.}}$ tends to be higher for a larger value of the ratio.
It is strongly suggestive that the tail in the weight distribution helps the QERC to acquire the generalization performance suppressing the training accuracy rate.
One can also find that the physical models (DTC and DDTC) have even smaller gaps.
Therefore, from our analysis, the DDTC model gives the best generalization performance and a nearly-optimal testing accuracy rate among quantum feature maps we have shown.
It is an encouraging fact that a simple Hamiltonian system could perform at least as good as a t-designed unitary map, which makes the implementation of such QNNs much simpler and more feasible.

\subsection{Simulations in other settings}

First, we consider the optimizer for the ONN.
We used the stochastic gradient descent method as the optimizer for the ONN in these numerical simulations.
The method has been broadly employed in many situations.
Hence, our observations above can be seen in many scenarios.
Moreover, we found the same effect of the tail with a more technical optimizer, AdaGrad \cite{Duchi2011} (see Appendix.~\ref{appendix_adagrad}).

Next, we direct our attention to the dataset.
So far, we have used the MNIST dataset to benchmark quantum feature maps and concluded that the tail in the weight distribution contributes to the testing accuracy rate and generalization performance.
In this section, we see the difference in the QERC performance between the quantum feature models we have considered with another dataset.
In order to see the difference in the generalization performance, we need to choose a dataset carefully.
If it's a simple dataset such as the 2D isotropic Gaussian samples demonstrated in Ref.~\cite{Sakurai2022}, all the feature models would have high accuracy rates, and it should be hard to see the difference between our models.
In contrast, if we choose a hard dataset such as the Fashion MNIST \cite{FashionMNIST}, all the models we have may achieve poor performance, and there should not be any room to discuss generalization.

For our calculations, we choose the Fashion MNIST dataset with a few classes since classifying all the classes is too hard for the QERC.
Then we picked three classes: T-shirt, Pullover, and Dress, and show the accuracy rate in Table.~\ref{acc_table}b).
One can see the same trend we have obtained with the MNIST dataset: the models with the tail tend to have higher testing accuracy rates and better generalization.
This examination suggests that the choice of a tailed feature map is an effective technique to push the QERC performance up more when the performance is good but not perfect.

\section{Discussion and Conclusion}
\label{sectionV}

In this work, we have performed network analysis on the unitary maps used in the QERC.  Such unitary maps $U$ can be converted to a weighted Hermitian matrix $G = i\log U$ that is characterized by the set of three weighted distribution functions. We observed that the weight distribution for the DTC model grows in time to near $n=50$ where it converges to its typical shape.  We compared the DTC model weight distribution against those associated with the Haar-random and Cauchy-random models. The DTC and Haar-random models are similar with respect to the Gaussian-like broadness of their weight distributions, whereas the DTC and  Cauchy-random models are similar in terms of the tails in their respective weight distribution. This suggests that the unitary map for DTC's period over $n=50$ is nearly as complex as the one by the Haar-random model, yet it still has power law characteristics and so is not totally random. 

Next, in the comparison of the performance of the QERC with the MNIST and Fashion MNIST datasets, we found that the power law tendency (tail) in the unitary map contributes to the high testing accuracy rate by suppressing the training accuracy rate.  This indicates that, at least for certain image classification problems, the tendency in the feature map in QNNs would help the QNNs to acquire better generalization performance.  Although similar observations have been noted in classical neural network models, including reservoir computation \cite{Zhang2018, Bertschinger2004, Boedecker2012, Toyoizumi2020}, this is the first time for it to be observed in the QNN scenario.

Not only the properties found here could serve as a guideline for designing more effective feature maps in the future, but our network approach to the quantum machine learning model could also be used to investigate other useful network properties in quantum feature spaces.
Furthermore, the approach could provide a physical interpretation of information processing in quantum machine learning schemes.
As the long tail in the weight distribution implies strong connectivities between certain basis states in the network dynamics, our approach may provide physical insights in the performance of quantum machine learning.

Finally, the fact that the QERC can perform with the feature map generated by a simple Hamiltonian model, such as the DTC with the disorder, is encouraging for the QERC's implementation. It strongly suggests that it can significantly reduce the overhead for the feature map in many other QNNs.

On the other hand, another overhead seems to remain; the reconstruction of the probability amplitudes of all computational basis states may require an exponentially large number of samples of the same initial states.
However, it is non-trivial whether such a precise reconstruction is needed for the ONN.
In fact, we need to acquire enough information for the task performed for the ONN.
Hence, it remains an open question how much it is possible to reduce the measurement overhead while maintaining the good performance with a simple Hamiltonian model used for the quantum feature map.

\appendix 

\section{Plot with error bars associated with Table.~\ref{acc_table}}\label{appendix_accplot}
\begin{figure}[t]
\centering
\includegraphics[width=9.2cm]{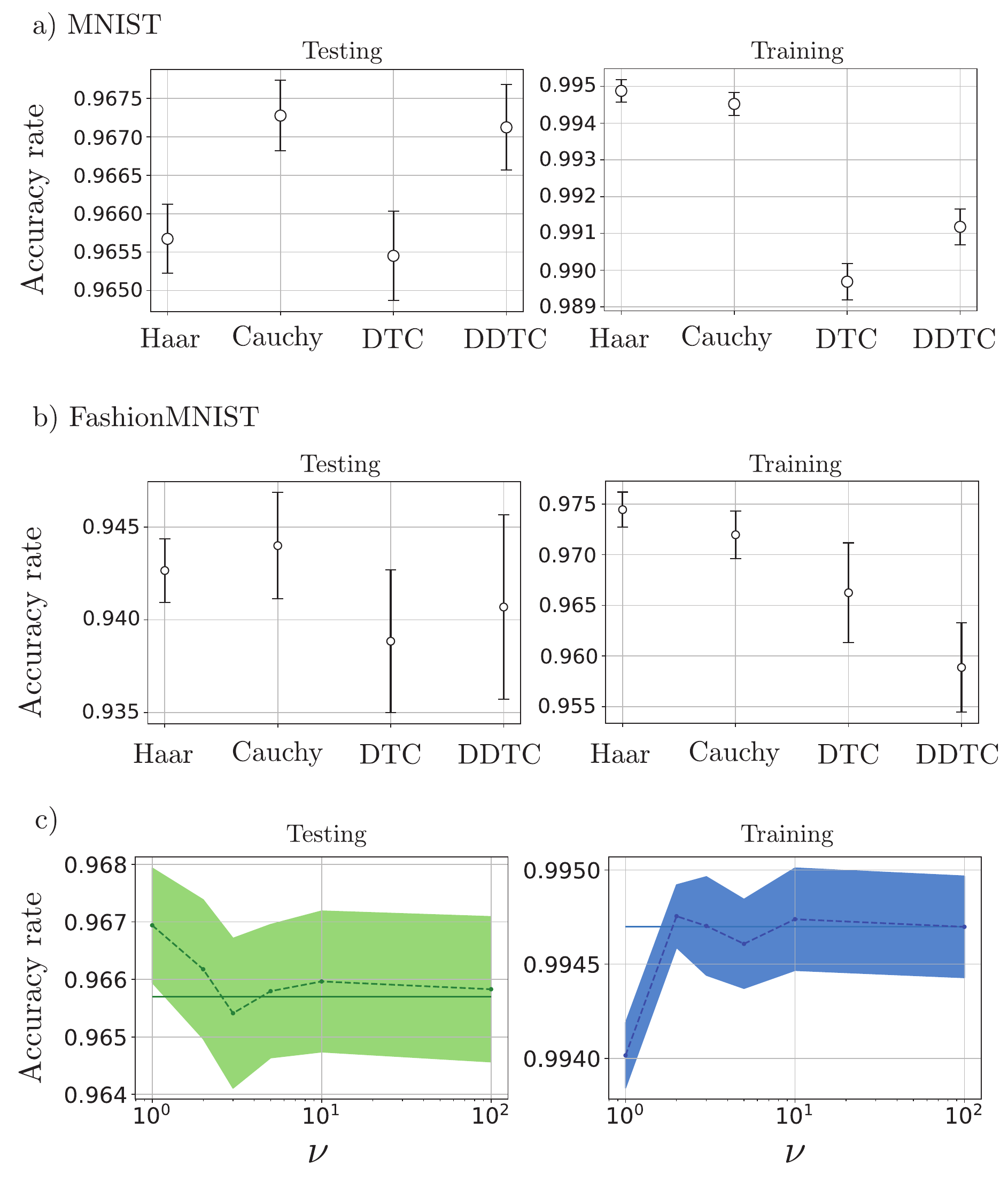}
\caption{Plot of the accuracy rates (with error bars) shown in Table.~\ref{acc_table} with a) the MNIST and b) FashionMNIST datasets. c) Accuracy rates for testing and training of the t-random model against $\nu$ with the MNIST dataset. The dashed lines and shaded areas correspond to the accuracy rates averaged over ten realizations for each value of $\nu$. The horizontal solid lines are the averaged accuracy rates of the Haar-random model.}\label{acc_error}
\end{figure}

Fig.~\ref{acc_error} is a plot associated with Table.~\ref{acc_table}.
It clearly shows the difference in the training and testing accuracy rates between different models.
In panel a), the improvements in testing by the tail are significantly observed.
In panel b), one may think that the effect of the tail is not so significant.
However, it is noteworthy that the training accuracy rate is significantly low in the Cauchy and DDTC models compared to the Haar and DTC models, respectively, whereas the testing accuracy rate is slightly higher.
Moreover, the error bars are associated with epochs, and we usually pick the neural network parameters at a single epoch with good performance.
Hence, we can extract the improvement by the tail in such a practice situation in the FashionMNIST case.

\section{Performance of the t-random model}\label{appendix_t}
Fig.~\ref{acc_error}c) shows the accuracy rates for training and testing against $\nu$.
The dashed lines and shaded areas correspond to the accuracy rates averaged over ten realizations for each value of $\nu$ and the associated standard deviations, respectively.
The horizontal solid lines are the averaged accuracy rates of the Haar-random model.
As $\nu$ increases, that is, as the effect of the tail decreases, the accuracy rates for both training and testing converge to the averaged one of the Haar-random model.

\section{Performance with the AdaGrad optimizer}\label{appendix_adagrad}

\begin{figure}[t]
\centering
\includegraphics[width=6cm]{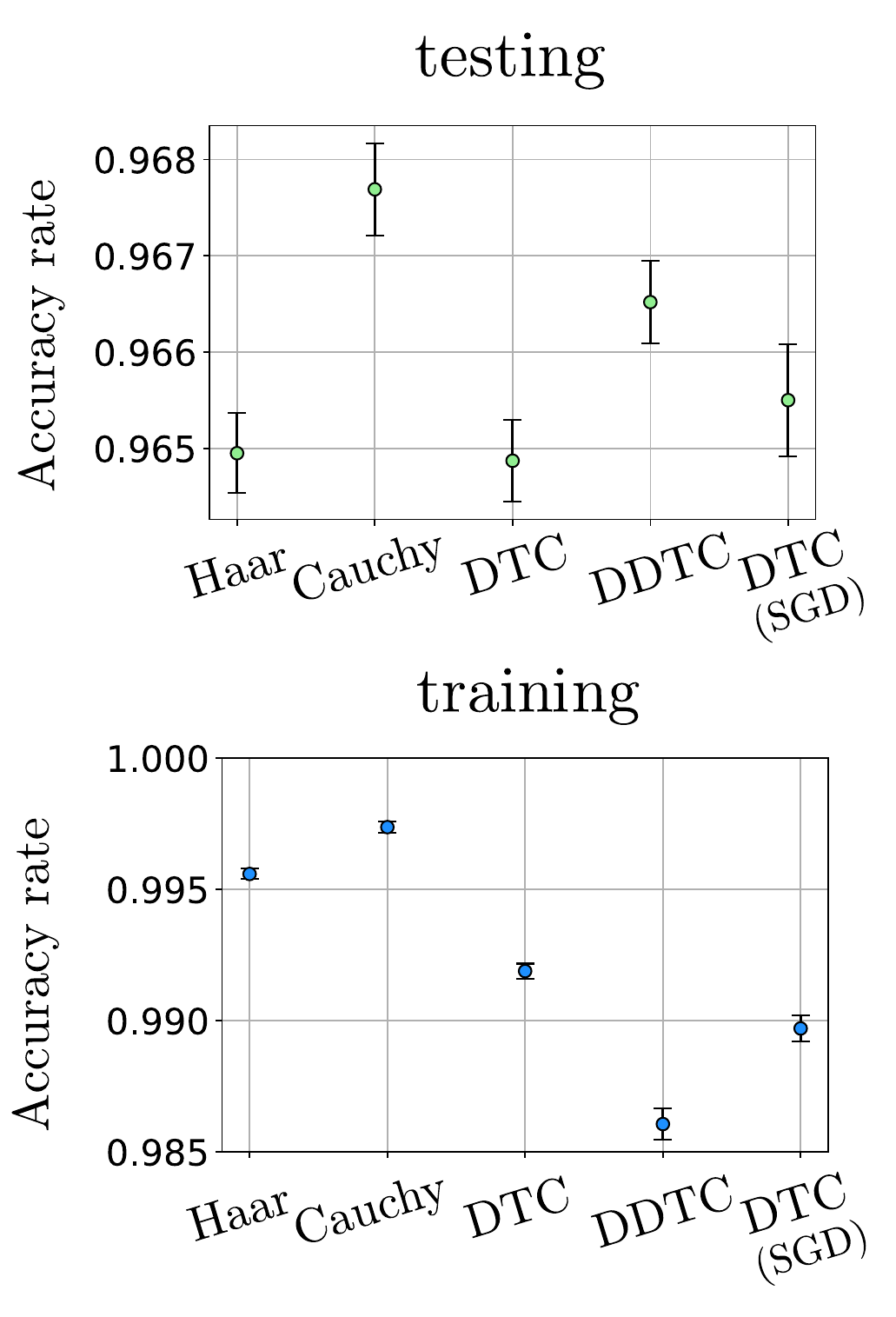}
\caption{Accuracy rates for testing and training of various models with AdaGrad. The same specific realizations are used for the random models as in the numerical simulations in the text. At each data point, the average is taken from 250 to 300 epochs. The error bar indicates the associated standard deviation.}\label{acc_error_AdaGrad}
\end{figure}

In this section, we will see the QERC performance with another optimizer for the ONN instead of the stochastic gradient descent (SGD) we have used in the text.
Here, we adopt AdaGrad \cite{Duchi2011} as the optimizer.
It uses an adaptive algorithm to update the parameters in the ONN based on the geometry of the data observed in earlier iteration steps.
Hence, it is considered that the training loss converges more quickly than with SGD.
In Fig.~\ref{acc_error_AdaGrad}, the accuracy rates for training and testing are shown for various models.
The accuracy rates for the DTC model with SGD are also plotted for comparison.
In fact, the training accuracy rate tends to be higher than that of the SGD case, and consequently, the generalization is poorer.
In this case, it can still be observed that the tail affects the testing accuracy rate and the generalization performance.
Furthermore, the DDTC model has a nearly-optimal testing accuracy rate and the best generalization in the AdaGrad case as well as the SGD case.

\section{Realization-averaged accuracy rates}\label{appendix_randomness}

\begin{figure}[ht]
\centering
\includegraphics[width=8.5cm]{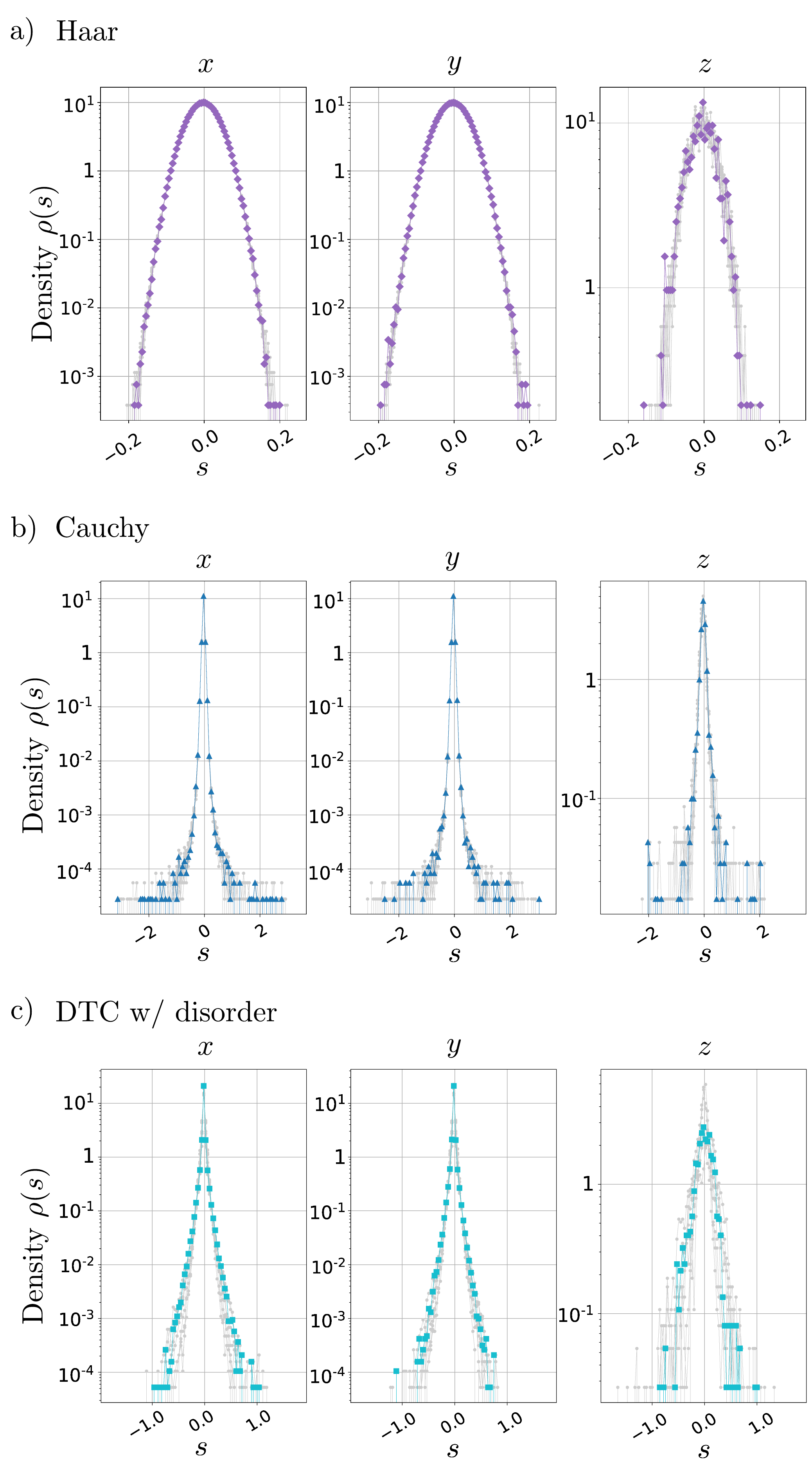}
\caption{Weight distributions of the Haar random model a), the Cauchy random model b), and the disordered DTC model c). The total number of realizations in each case is 10. The colored curves are corresponding to each curve in Fig.~\ref{Comparison_distributions} a,b,d) respectively. The other realizations are plotted as gray curves.}\label{realization}
\end{figure}

In Fig.~\ref{realization}, we showed that the weight distribution of each model is depicted with ten realizations. The colored curves here correspond to those in Fig.~\ref{Comparison_distributions} a,b,c). We clearly observe that the width and tail properties of the weight distribution are not strongly dependent on the particular realizations.

The realization-averaged accuracy rates for the Haar-random, Cauchy-random, and disordered DTC models are also shown in Table.~\ref{avg_acc_table}. As the mean difference $\Delta_\text{acc.}$ shows, the models which have a tail in the weight distribution potentially avoid overlearning. The disordered DTC model achieves the highest generalization performance.

\begin{table}[t]
\centering
\resizebox{\columnwidth}{!}{%
\begin{tabular}{|c||c|c|c|}
\hline
MNIST & testing (std.)& training (std.) & $\Delta_\text{acc.}$ \\ \hline\hline
Haar & 0.9657 ($\pm 0.0011$) & 0.9947 ($\pm 0.0003$)& 0.0290 \\ \hline
Cauchy & 0.9678 ($\pm 0.0013$) & 0.9943 ($\pm 0.0002$) & 0.0265 \\ \hline
DTC w/ disorder & 0.9668 ($\pm 0.0005$) & 0.9910 ($\pm 0.0002$) & 0.0242\\ \hline
\end{tabular}}
\caption{Realization- and epoch-averaged accuracy rates of the various models. The epoch average involves from 250 to 300 epochs. The realization average is taken over 10 realizations for each model with the means (and standard deviations in parenthesis) shown. $\Delta_\text{acc.}$ denotes the difference between the means for training and testing.}
\label{avg_acc_table}
\end{table}

\acknowledgements{We thank Victor M. Bastidas for valuable discussions. This work is supported by the MEXT Quantum Leap Flagship Program (MEXT Q-LEAP) under Grant No. JPMXS0118069605 and JST the establishment of university fellowships towards the creation of science technology innovation under Grant Number JPMJFS2136.}

\newpage

%

\end{document}